\documentclass[reprint,amsmath,amssymb,aps]{revtex4-2}

\usepackage{graphicx}
\usepackage{dcolumn}
\usepackage{bm}
\usepackage{harpoon}

\usepackage{makecell}
\usepackage[colorlinks,
            linkcolor=red,
            anchorcolor=blue,
            citecolor=blue
            ]{hyperref}

\begin{document}

\preprint{APS/123-QED}

\title{New Experimental Limits on Exotic Spin- and Velocity-dependent Interactions Using Rotationally Modulated Source-masses and an Atomic-magnetometer Array}

\author{K.Y. Wu}
\affiliation{Key Laboratory of Neutron Physics, Institute of Nuclear Physics and Chemistry, CAEP, Mianyang, Sichuan, 621900,China }
\affiliation{Institute of Nuclear Physics and Chemistry,CAEP, Mianyang, Sichuan, 621900,China }

\author{S.Y. Chen}
\affiliation{Key Laboratory of Neutron Physics, Institute of Nuclear Physics and Chemistry, CAEP, Mianyang, Sichuan, 621900,China }
\affiliation{Institute of Nuclear Physics and Chemistry,CAEP, Mianyang, Sichuan, 621900,China }

\author{G.A.Sun}
\affiliation{Key Laboratory of Neutron Physics, Institute of Nuclear Physics and Chemistry, CAEP, Mianyang, Sichuan, 621900,China }
\affiliation{Institute of Nuclear Physics and Chemistry,CAEP, Mianyang, Sichuan, 621900,China }

\author{S.M. Peng}
\affiliation{Institute of Nuclear Physics and Chemistry,CAEP, Mianyang, Sichuan, 621900,China }

\author{M. Peng}
\affiliation{Key Laboratory of Neutron Physics, Institute of Nuclear Physics and Chemistry, CAEP, Mianyang, Sichuan, 621900,China }
\affiliation{Institute of Nuclear Physics and Chemistry,CAEP, Mianyang, Sichuan, 621900,China }

\author{H.Yan}
\email[Corresponding author: ]{hyan@caep.cn}\affiliation{Key Laboratory of Neutron Physics, Institute of Nuclear Physics and Chemistry, CAEP, Mianyang, Sichuan, 621900,China }
\affiliation{Institute of Nuclear Physics and Chemistry,CAEP, Mianyang, Sichuan, 621900,China }
%
%
%

\date{\today}

\begin{abstract}
 We conducted laboratory searching for the exotic spin- and velocity-dependent new interactions according to the previously proposed experimental scheme. Two $\sim$6Kg heavy source masses are rotationally modulated at a frequency of 20Hz. Four identical atomic magnetometers are used in an array form to increase the statistics and cancel the common-mode noise.
Data processing method based on high precision numerical integration is applied for the four harmonic frequencies of the signal. The rotation direction of the source masses was reversed to flip the signal. Thus the [1,-3,3,-1] weighting method can be applied to remove possible slow drifting further. 
The experiment method has noise reduction features, and new constraints for Vector-Axial and Axial-Axial were obtained. The new constraints on VA improved by as much as more than four orders, on AA by as much as two orders in the corresponding force range, respectively.
\end{abstract}

\maketitle

\section{\label{sec:level1}Introduction}
New physics beyond the Standard Model is possible. New interactions mediated by new particles are solutions to several important questions in modern physics, such as the strong CP problem\cite{PIE2021} and the dark matter\cite{PDG2020}. Long ago, Peccei and Quinn\cite{PEC77} proposed the PQ mechanism to solve the strong CP problem. Wilczek\cite{WIL78} and Weinberg\cite{WEI78} simultaneously noticed the PQ mechanism would generate new pseudo-scalar particles, which are named axions. In 1984, Moody and Wilczek~\cite{1984New}pointed out that ALPs(axion-like particles) could mediate macroscopic spin-dependent interactions. Many experiments have been performed to search for and constrain the scalar-pseudoscalar type interaction. Later on, start from rotational invariance, Dobrescu and Mocioiu~\cite{Dobrescu2006} formed 16 different operator structures involving the spin and momenta of the interacting particles. ALPs mediated new interactions are a subset of the new theory. Now the force carriers could also be vector particles. As early as 1980, Fayet\cite{FAY1980a,FAY1980b} pointed out that the new U(1) vector bosons with small masses and weak couplings to ordinary matter can be produced by spontaneously breaking of the supersymmetric theories. Searching for the new interactions mediated by the new particles is related to the strong CP problem, dark matter, and supersymmetry, which are the most important unsolved problems in modern physics.

For the vector force carriers, the interaction can be deduced from the coupling $\mathcal{L}_{X}=\bar{\psi}(g_{V}\gamma^{\mu}+g_{A}\gamma^{\mu}\gamma_{5})\psi X_{\mu}$ where $X_\mu$ is the new vector particle. There are  the VA(vector-axial-vector) interaction $V_{VA}(r)$($V_{12,13}$ in Ref.~\cite{Dobrescu2006}'s notation) and AA(axial-axial) interaction $V_{AA}(r)$($V_{4,5}$ in Ref.~\cite{Dobrescu2006}'s notation),:
  \begin{eqnarray}\label{eqnVA}
  V_{VA}(r)=\frac{\hbar g_{V}g_{A}}{2\pi}\frac{\exp{(-r/\lambda)}}{r}\vec{\sigma}\cdot\vec{v}\\
  V_{AA}(r)=\frac{\hbar^{2}g_{A}^{2}}{16\pi mc}(\frac{1}{\lambda r}+\frac{1}{r^{2}}) {\exp{(-r/\lambda)}}\vec{\sigma}\cdot(\vec{v}\times\hat{r})
  \end{eqnarray}
where $\vec{v}$ is the relative velocity between the probe particle and source particle,$\lambda=\hbar/m_{X}c$ is the interaction range, $m_{X}$ is the mass of the new vector boson, m the mass and $\vec{\sigma}$ the Pauli matrices of the spin-polarized probing particle. $g_Vg_A$ and $g_Ag_A$ are the interaction coupling constants that are both dimensionless.

ALPs(Axion Like Particle) are very difficult to search for in the laboratory, and they have eluded detection so far. Since these new light bosons can mediate macroscopic interactions, one method to search for these new particles is to probe the boson field caused by a macroscopic body.  There are typically two methods to look for the new interactions, to detect the macroscopic forces or the torques exerted on the polarized probe spins. For example, Leslie et al. \cite{LES2014} proposed experimental schemes to detect the new spin-dependent force between the spin-polarized source and a mechanical oscillator. For another example, Ding et al. used a micro-fabricated magnetic structure as the polarized source, then tried to detect the AA type force in a range of $\sim\mu$m sensed by a gold-sphere-cantilever~\cite{Ding2020}. Many people search for the new interaction through its rotating effects as a pseudo-magnetic field on the polarized spin. The VA and AA interactions between different combinations of Fermions had been investigated already, such as electron-nucleon ~\cite{Kim2018,Kim2019,Ji2018,Ding2020}, neutron-nucleon ~\cite{Piegsa2012,Yan2013,Yan2015}, electron-electron ~\cite{Ficek2017}, electrons and antiprotons ~\cite{Ficek2018}, electron-nucleon~\cite{Yan2013}. Studies on these new interactions involving muons were performed very recently\cite{YAN19}. 
 Ref.\cite{Safronova2018} is an extensive review of recent theoretical and experimental progress of the new interaction searching research.In this work, we are interested in detecting new interactions by measuring the pseudo-magnetic field caused by the source mass to the polarized electron spin. 

Due to its high sensitivity based on polarized electron spins\cite{BUD2013}, AMs(Atomic Magnetometer) is convenient for searching these exotic spin-dependent new interactions. Kim et al. employed a commercial cm-scale AM to detect the interaction between polarized valence electrons of Rb in the vapor cell of magnetometer and nucleons~\cite{Kim2018,Kim2019}.  
 The commercially available AMs have relatively lower sensitivities but are compact and can be easily arranged in an array of 50 units to measure the bio-magnetic field generated by the human brain\cite{BOT2021,REA2021}. To the best of our knowledge, the AMs in the array form has never been used to search for spin-dependent new interactions.
  
 In Ref.\cite{WU2021}, we proposed to use rotationally-modulated source masses and an array of AMs to search for the exotic spin-dependent interactions. Monte Carlo simulations indicate that the new experiment scheme can achieve sensitivity improvement as much as $\sim$5 orders.
 We follow the previously proposed scheme to search for the VA and AA type new interactions in this work.
 
\section{\label{sec:level1}The Basic Idea}
We first briefly review the basic idea of the previously proposed experiment scheme\cite{WU2021}. Two dense, identical source masses are rotationally modulated with frequency $f_0$. An array consisted of four identical AMs is placed symmetrically around the source masses. If the exotic spin-dependent interactions exist, a pseudo-magnetic field can be induced by the source masses. Theoretically, the pseudo magnetic field at the point $\vec{r}$ can be calculated and expressed as:\begin{eqnarray*}
\label{B'}
\vec{B'}_{VA}(\vec{r}) =\frac{ g_{V}g_{A}}{\pi\gamma_e}\int d^3\vec{r}'\frac{\exp{(-|\vec{r}-\vec{r}'|/\lambda)}}{|\vec{r}-\vec{r}'|}\vec{v}\\
 \vec{B'}_{AA}(\vec{r})=\frac{\hbar g_{A}^{2}}{8\pi m_ec\gamma_e}\int d^3\vec{r}'(\frac{1}{\lambda |\vec{r}-\vec{r}'|}+\frac{1}{|\vec{r}-\vec{r}'|^{2}})\times\\
 {\exp{(-|\vec{r}-\vec{r}'|/\lambda)}}(\vec{v}\times\frac{\vec{r}-\vec{r}'}{|\vec{r}-\vec{r}'|})
\end{eqnarray*}
 where  $\gamma_e$ is the gyromagnetic ratio of the electron, $d^3\vec{r}'$ is a three-dimensional volume element at $\vec{r}'$ of the source mass. Now the probing polarized particle is assumed to be the electron since the AM is using polarized electrons. Taking $g_Vg_A=g_A^2=1$, at the position of the AM, along the most significant direction, the pseudo magnetic field induced by the rotating source masses can be expanded in Fourier series as:
  \begin{eqnarray*}
\label{B'}
B'(t) = {c_0} + \sum_{n=1}^{\infty}{c_n}\cos ({n\omega _0}t+\phi) 
 \end{eqnarray*} 
where $c_0$ is the DC component, $\phi$ the initial phase factor and $c_n$ can be expressed as:
\begin{eqnarray}
\label{cn}
{c_n} 
=\frac{2}{NT}\int_0^{NT}  \cos \left( {n{\omega _0}t+\phi} \right){B'(t)}dt
\end{eqnarray}
where $N$ is an integer, $T=1/f_0=2\pi/\omega_0$ the period of the modulated signal, thus $NT$ is the total observing time. In principle, $B'(t)$ can be calculated using Monte Carlo techniques and $c_n$ using numerical integration methods.
 In the experiment, when taking into account the noise, the detected signal is:
\begin{eqnarray}
\nonumber {B_{\exp }(t)} &=& \alpha{c_0} + \alpha\sum_{n=1}^{\infty}{c_n}\cos ({n\omega _0}t+\phi)  + n(t)
 \end{eqnarray}
 where $\alpha=g_Vg_A$ for the VA interaction and $\alpha=g_A^2$ for the AA interaction respectively, and $n(t)$ the noise.
For each harmonic component, $\alpha$ thus the coupling constant can be derived as:
 \begin{equation}
\alpha|_n= \frac{{2\int_0^{NT}  {\cos \left( {n{\omega _0}t+\phi} \right){B_{exp}(t)}dt} }}{{c_n{NT}  }}
\end{equation}
According to the Fourier expansion of $B'(t)$,  harmonic terms of one to four using weighted mean can be used to determine $\alpha$,
\begin{eqnarray}
\bar{\alpha}= \frac{{\sum\limits_{n = 1}^4 {c_n^2{\alpha|_n}} }}{{\sum\limits_{n = 1}^4 {c_n^2} }}
\end{eqnarray}
 The noise contribution can be estimated as\cite{LIB2003}:
\begin{equation}\label{eqn.lok}
\delta\bar{\alpha}|\sim \sqrt{S_N(nf_0)}\sqrt{\frac{2}{NT}}\frac{1}{\sqrt{\sum\limits_{n = 1}^4 {c_n^2}}}
\end{equation}
 where $S_N(nf_0)$ is the noise power density at the corresponding harmonic frequency. In this work, $S_N(nf_0)$ is found close to be a constant at a level of $\sim10fT/\sqrt{Hz}$ for the interested harmonic frequencies. The DC term is not used to avoid the $1/f$ noise of the AMs. The integration method applied on the harmonic terms has the feature of reducing noise bandwidth. The most significant harmonics are used in the weighted average way could further reduce the noise, as it can be seen from Eqn.(\ref{eqn.lok}).
  
Furthermore, four identical AMs arranged in an array are used to improve the statistics and reduce the common-mode noise. More details will be described later. 
 \section{\label{sec:level1}The Experimental Setup}

\begin{figure*}
\centering
\includegraphics[width=17cm]{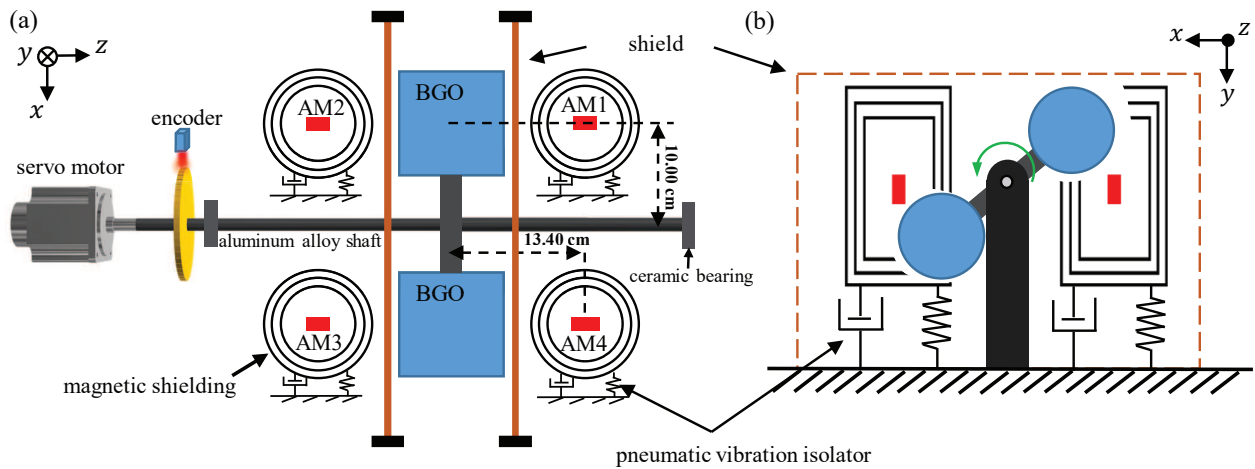}
\caption{\label{fig:setup}Schematic of the experimental setup. (a) and (b) are the top and end views of the setup, respectively. The servo motor rotates the two BGO cylinders as source masses with the modulating frequency of 20Hz, inducing effective magnetic field signals to the surrounding AMs if exotic spin- and velocity-dependent interactions exist. The AMs are magnetically shielded and placed on a platform sitting on pneumatic vibration isolators. A 1cm thick aluminum plate shielding is applied to prevent possible air vibration caused by the rotating source masses. The encoder monitors the rotating angle and frequency in real-time.}
\end{figure*}

\begin{figure}[htp]
\includegraphics[width=8cm]{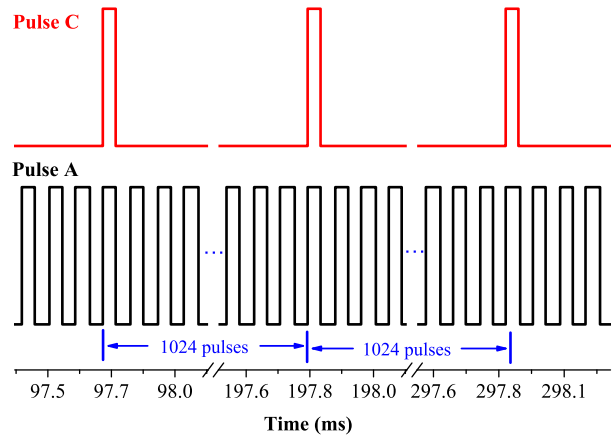}
\caption{\label{fig:PAPC}Pulse examples of Phase A and C signals of the 3-phase encoder.}
\end{figure}

The experimental setup is shown as FIG.~\ref{fig:setup}. Two large, identical BGO (Bi$_4$Ge$_3$O$_{12}$) crystal cylinders are attached to the rotating shaft of a high-power servo motor. The BGO crystals have high purity(99.9999\% ), high mass density(7.13g/$cm^3$), and very low magnetic susceptibility\cite{2003Scintillator}. All these features make the crystal to be a good choice as the source masses for searching the exotic spin-dependent interactions\cite{TUL2013,Kim2018,Kim2019}. Two BGO crystal cylinders with a length of 10.16cm and a diameter of 10.16cm are attached to the shaft of the servo motor. The mass of a single BGO crystal is 5.87Kg. The shaft material is chosen to be the aluminum alloy of 7475, which is strong and has a lower magnetic susceptibility than the common 6xxx series alloys. We also carefully chose the ceramic bearings to support the rotating shaft. The servo motor rotates the two crystals at a frequency of $\sim$10Hz; thus, the source masses are rotationally modulated at a frequency of $\sim$20Hz. A fiber-optic-encoder is applied to monitor the precise rotating angle and frequency in real-time. The 3-phase encoder has three output signals which are called phase A, B, and C. Phase A(FIG.\ref{fig:PAPC}) and B signals are pulses with a 90$^{\circ}$ phase difference, and they give the rotational direction and angle in time series. Phase C signal indicates the particular reference angle of the encoder, and it gives one pulse per rotation as shown in FIG.\ref{fig:PAPC}.

\begin{figure}[h]
\includegraphics[width=8.5cm]{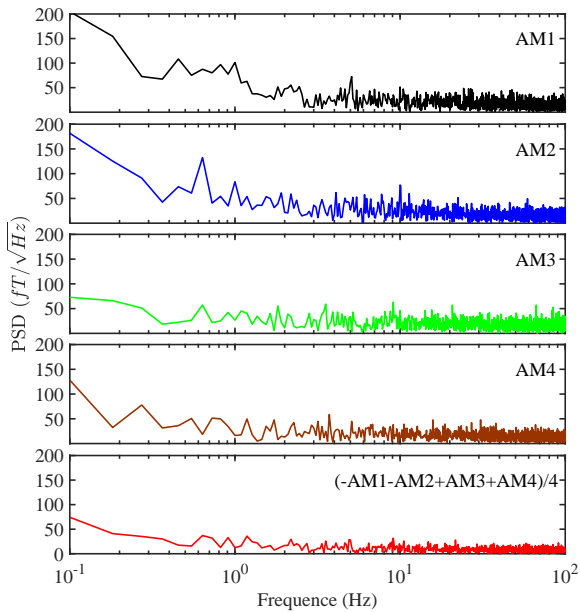}
\caption{\label{fig:PSD1}Noise-power-density measurement examples of the VA searching setup. The 10Hz vibration noise can be seen on AM1 and AM2 but not on AM3 and AM4. As shown in the bottom figure, common-mode-noise is reduced by using Eqn(7), especially for the low-frequency cases.  }
\end{figure}

An array of four identical, high sensitivity, commercially available AMs is used to detect the new interactions. Each AM is placed inside a magnetic shielding which has an outer diameter of 12cm and length of 28cm and is made of 3 layers of permalloy. The magnetic shielding provides a magnetic environment of residual field level of $\sim$5nT; thus, the AMs can work at their best sensitivities. The magnetic shielding can screen possible magnetic noise, which will disturb the polarized electron spin inside the shielding while the new spin-dependent interactions will not be affected. The AMs have a bandwidth of 200Hz, which is good enough since the highest harmonic frequency under consideration is $\sim$80Hz. The exotic spin-dependent interactions due to the source masses, if they exist, can induce pseudo magnetic field signals for the polarized electron spin of the surrounding AMs. 
The AMs are the dual-axis and can be set to be sensitive to either $\hat{x}$ or $\hat{y}$ direction in our setup as shown in FIG.\ref{fig:setup}. To detect the VA type interaction, the AMs are set to measure along $\hat{y}$ direction, and the signal due to the new interaction can be extracted as
\begin{equation}
\frac{1}{4}(-AM1-AM2+AM3+AM4)
\end{equation}
when the source masses are rotating clockwise(CW), obviously, any common noises that get added to the signals can be canceled. A data-taking example of this configuration is shown as FIG.\ref{fig:PSD1} which clearly shows the common-mode noise reduction feature of the experiment method.
When searching for the AA type new interaction, the AMs are set to be sensitive on $\hat{x}$ direction, and the new force signal can be extracted as:
\begin{equation}
\frac{1}{4}(-AM1+AM2-AM3+AM4)
\end{equation}
when rotating CW.   The same noise-canceling mechanism works as well.

Since two heavy source masses are rotating at a speed of 600RPM, one of the main issues of the experiment is to isolate the vibrations effectively. The AMs, cables, electronics, and other necessary parts are all placed on a pneumatic vibration-isolation platform with a resonance frequency of 1Hz. Although the 10Hz vibration can still be seen on some AMs as in FIG.\ref{fig:PSD1}, it shows no presence on the interested harmonic frequencies of 20, 40, 60, and 80Hz.
Furthermore, shielding made of the Aluminum plate with 1cm thickness is applied to prevent possible air vibration, which is also caused by the rotations of the heavy masses and might disturb the AMs. Closed cable trunking is used to shield all the necessary parts of the connecting cables. To further reduce the noise,  the AMs are  powered by a UPS, and the 50Hz peak of the line power is not seen as in our previous measurement(FIG.2 of Ref.\cite{WU2021}).

A digital, multichannel data acquisition card(DAQ) with a maximum sampling rate of 2MHz is used to read the output signals of the encoder and AMs synchronously. 
The data-taking cycles are as follows. The source masses were first rotated clockwise for 660s then counterclockwise for the same amount of time. Each 660s cycle is further divided into 60 segments of length 11s to avoid over stacking the DAQ card.
 When the rotation direction is reversed, the signal due to the new interaction changes its sign while the noise will not be affected. 
For the VA type interaction searching, the total data integration time is 130h, and for AA, 243h.

 \section{\label{sec:level1}Data processing and results}

The data process procedure is as follows. For each data segment of 11s, the rotating or the modulating frequency $f_0$ is obtained with a typical error of $\sim3\times10^{-4}$Hz by fitting the time series from the Phase A signal of the encoder. Then the period, T of the modulated signal can be calculated. The 11s data segment can be truncated to be an integer number of the period to avoid unnecessary uncertainties for performing the integrations. The initial phase of the system can be determined with a typical error of 0.024$^{\circ}$, using the Phase C signal from the encoder. With the known $\omega_0$ and $\phi$, $B'(t)$ can be calculated using Monte Carlo techniques as in Ref.\cite{Kim2018,Kim2019,Ji2018}. Then $c_n$ can be obtained by numerically integrating Eqn.(3). Once $c_n$ is obtained, $\alpha$ ,or $g_Vg_A$ and $g_Ag_A$ ,can be calculated by integration of Eqn.(4) using $B_{exp}(t)$ 
time-series measured by the AMs. Simpson's method, which is a numerical integration technique with high precision\cite{YAN2014}, is applied throughout the work. Once $\alpha$s and $-\alpha$s are obtained for each CW and CCW cycle, the [+1,-3,+3,-1] weighting method\cite{Kim2018,Kim2019,Chu2013PRD} is applied to remove possible slow drifting of the system further. 

 \begin{figure}[h]
\includegraphics[width=8.5cm]{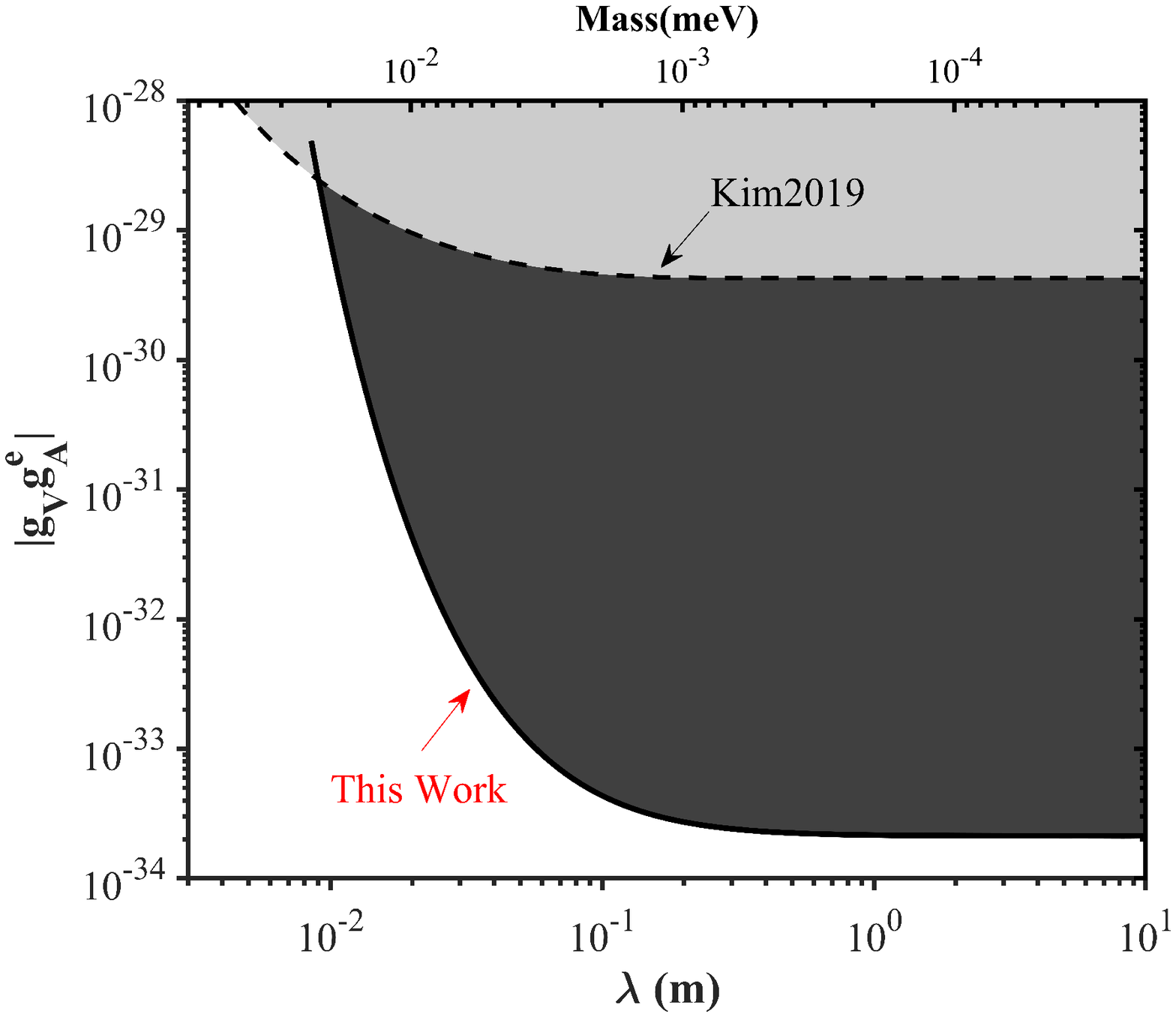}
\caption{\label{fig:aVA}Constraints on the coupling constants $|g_Vg_A^e|$ (1$\sigma$) as a function of force range $\lambda$ and new boson mass. The dashed line is from Ref~\cite{Kim2019}. The solid line (dark area) is the present work. }
\end{figure}
 \begin{figure}[h]
\includegraphics[width=8.5cm]{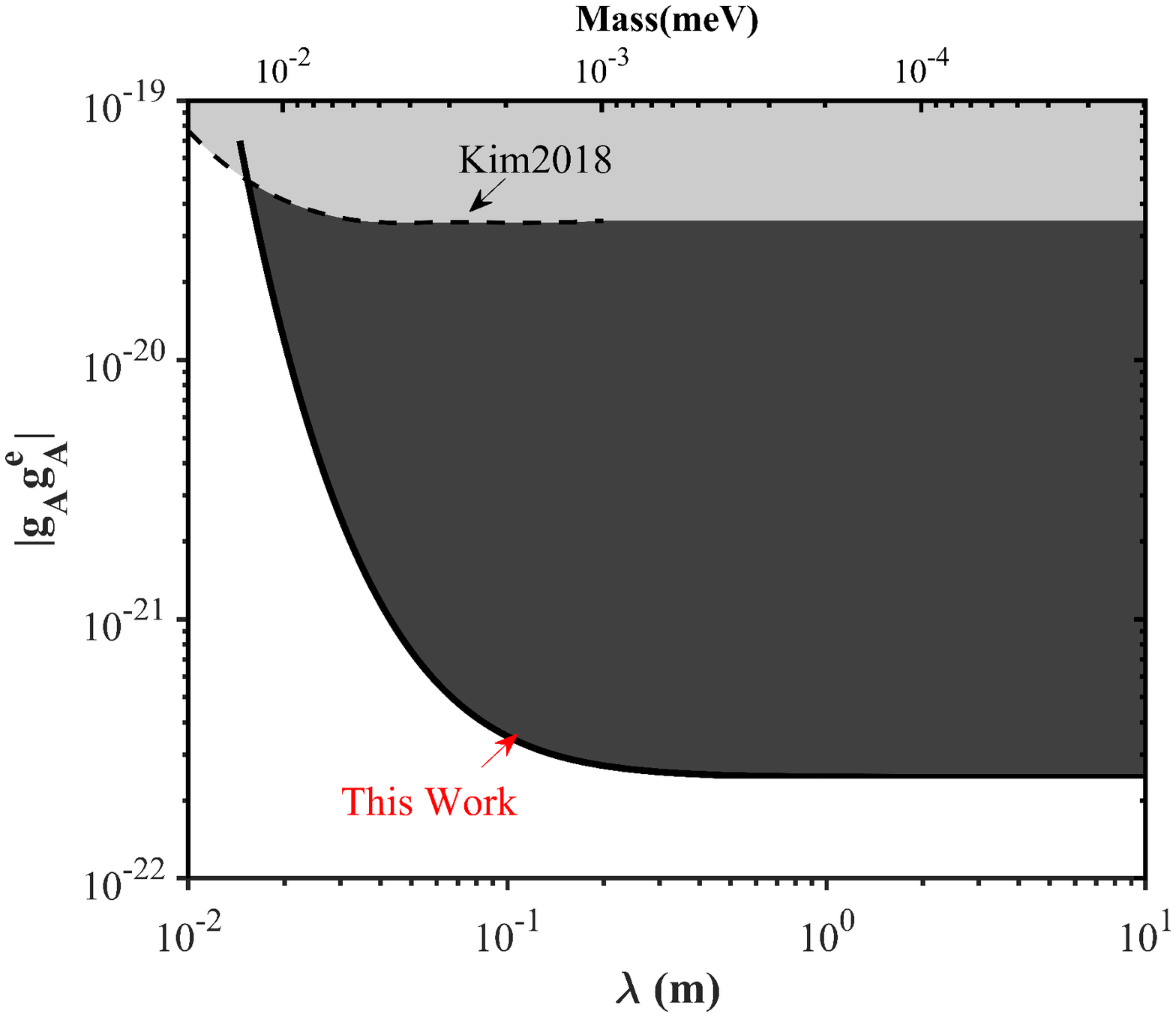}
\caption{\label{fig:aAA}Constraints on the coupling constants $|g_Ag_A^e|$ (1$\sigma$) as a function of force range $\lambda$ and new boson mass. The dashed line is from Ref~\cite{Kim2018}. The solid line (dark area) is the present work. }
\end{figure}

Errors due to uncertainties of $f_0$, $\phi$, rotation radius, the distance between AMs and the source masses, contributions from the rotating aluminum parts, etc., were carefully analyzed. The systematic errors of the experiment were found to be well below  the statistics. For $\lambda=10$m, we observed:
\begin{eqnarray}
g_Vg_A^e=0.07\pm2.06\times10^{-34}\\
g_Ag_A^e=-0.06\pm2.36\times10^{-22}
\end{eqnarray}

FIG.\ref{fig:aVA} and FIG.\ref{fig:aAA} are the main results of this work. FIG.\ref{fig:aVA} shows our upper limit on the VA coupling strength as a function of the range $\lambda$ and the new boson mass. The new result excludes the dark gray area in the $(g_Vg_A^e,\lambda)$ plane. The limits shown involve the axial coupling $g_A^e$ of the electron and the vector coupling $g_V$ of an unpolarized ensemble composed of nucleons and electrons. Our bounds $|g_Vg_A^e|<10^{-29}$ for $\lambda=1$cm to $|g_Ag_A^e|<2.1\times10^{-34}$ for $\lambda=10$m are the most stringent laboratory limits over these distances. FIG.\ref{fig:aAA} shows our upper limit on the AA coupling strength as a function of the range $\lambda$ and the new boson mass. The new result excludes the dark gray area in the $(g_Ag_A^e,\lambda)$ plane. The limits shown involve the axial coupling $g_A^e$ of the electron and the vector coupling $g_A$ of the source mass. Our bounds $|g_Ag_A^e|< 1.2\times10^{-20}$  for $\lambda=2$cm to $|g_Ag_A^e|<2.4\times10^{-22}$  for $\lambda=10$m are the most stringent laboratory limits over these distances.

 \section{\label{sec:level1}Conclusion and Discussion}
The experimental scheme, searching for VA and AA type new interactions proposed previously, was realized in this work. FIG.\ref{fig:aVA} and \ref{fig:aAA} are the main results of the present work.  In the force range from $\sim$0.2m to 10m, the present results improve the limits on $|g_Vg_A^e|$ by $\sim$20000 times and on  $|g_Ag_A^e|$ by  $\sim$140 times.  The major systematic problem for the experiment is the vibrations caused by the high-speed rotating of the heavy source masses. We successfully circumvent it by using the pneumatic vibration isolation techniques.  At this stage, it is not clear to us how to further improve the present experiment method. We noticed that method of spin amplifiers\cite{SU2021} had been proposed recently. Based on that scheme, polarized $^3$He can be advantageous for having a high amplification factor. We could conceive that if polarized $^3$He cells can be arranged in an array form as in this work, better sensitivity might be achieved. 

We acknowledge support from the National Key Program for Research and Development of China under grant 2020YFA0406001 and 2020YFA0406002. This work was also supported by the National Natural Science Foundation of China(Grant U2030209). We thank ChongQing Medical University for the loan of the AMs. We thank Dr. W.Ji, H.F.Dong, and H.Yuan for helpful discussions.


\end{document}